\documentclass[sigconf]{acmart}

\AtBeginDocument{%
  \providecommand\BibTeX{{%
    \normalfont B\kern-0.5em{\scshape i\kern-0.25em b}\kern-0.8em\TeX}}}

\setcopyright{acmcopyright}
\copyrightyear{2018}
\acmYear{2018}
\acmDOI{10.1145/1122445.1122456}

\acmConference[Woodstock '18]{Woodstock '18: ACM Symposium on Neural
  Gaze Detection}{June 03--05, 2018}{Woodstock, NY}
\acmBooktitle{Woodstock '18: ACM Symposium on Neural Gaze Detection,
  June 03--05, 2018, Woodstock, NY}
\acmPrice{15.00}
\acmISBN{978-1-4503-XXXX-X/18/06}

\usepackage{textgreek}
\usepackage[inline]{enumitem}
\usepackage{subcaption}
\usepackage{tikz}
\usetikzlibrary{calc}
\usetikzlibrary{positioning}
\usetikzlibrary{arrows}
\usepackage{adjustbox}
\usepackage{algpseudocode}
\usepackage{algorithm}
\usepackage{seqsplit}

\algdef{SE}[VARIABLES]{Variables}{EndVariables}
{\algorithmicvariables}
{\algorithmicend\ \algorithmicvariables}
\algnewcommand{\algorithmicvariables}{\textbf{global variables}}

\definecolor{mGreen}{rgb}{0,0.6,0}
\definecolor{mGray}{rgb}{0.5,0.5,0.5}
\definecolor{mPurple}{rgb}{0.58,0,0.82}
\definecolor{backgroundColour}{rgb}{0.99,0.99,0.99}

\newcommand{\ie}{\textit{i.e.,} }
\newcommand{\eg}{\textit{e.g.,} }
\newcommand{\wrt}{\textit{w.r.t.\ }}

\begin{document}

\title[$\text{SAFE}^d$: Self-Attestation For Networks of 
Heterogeneous Embedded Devices]{$\text{SAFE}^d$: Self-Attestation\\For Networks 
of Heterogeneous Embedded Devices}

\author{Alessandro Visintin}
\email{visintin@math.unipd.it}
\affiliation{%
  \institution{University of Padua}
  \country{Italy}
}

\author{Flavio Toffalini}
\email{flavio\textunderscore toffalini@mymail.sutd.edu.sg}
\affiliation{%
  \institution{Singapore University of Technology and Design}
  \country{Singapore}
}

\author{Mauro Conti}
\email{conti@math.unipd.it}
\affiliation{%
  \institution{University of Padua}
  \country{Italy}
}

\author{Jianying Zhou}
\email{jianying\textunderscore zhou@sutd.edu.sg}
\affiliation{%
  \institution{Singapore University of Technology and Design}
  \country{Singapore}
}


\renewcommand{\shortauthors}{Visintin, et al.}

\begin{abstract}	
The Internet of Things (IoT) is an emerging paradigm that 
allows to set large networks of small 
and independent devices.
To ensure their integrity, practitioners employ 
so-called Remote Attestation (RA) schemes.
Classic RA schemes require a central and powerful entity, 
called Verifier, that has mainly two duties: 
\begin{enumerate*}[label=(\roman*)]
	\item it manages the entire process of attestation, and
	\item it contains all the proofs for validating the devices' integrity.
\end{enumerate*}
However, having a central Verifier makes the 
network dependent upon an external entity and introduces a single point of 
failure for security.

In this work, we propose $\text{SAFE}^d$: the first RA schema
that allows a pair of IoT devices to validate their integrity
without relying on an external Verifier.
Our approach overcomes previous limitations by spreading the
proofs among multiple IoT devices and using novel 
cryptographic mechanisms to ensure secure communications.
Moreover, the entire IoT network can collaboratively isolate
tampered devices and recover missing proofs in case 
of anomalies.

We evaluate our schema through an implementation for 
Raspberry Pi platform and a network simulation.
The results show that $\text{SAFE}^d$ can detect infected
devices and recover up to $99.9\%$ of proofs in case of faults or 
attacks.
Moreover, we managed to protect up to $10$K devices with a logarithmic overhead 
on the network and on the devices' memory.

\end{abstract}



\maketitle

\section{Introduction}
Internet of Things (IoT) refers to a set of new technologies 
that allow building sophisticated applications by using 
groups of small and interconnected devices.
IoT world revolves around three cardinal concepts:
\begin{enumerate*}[label=(\roman*)]
	\item a decentralized network which can be accessed remotely through the Internet,
	\item heterogeneous devices that collaborate autonomously, and
	\item interaction with the physical environment via sensors and actuators.
\end{enumerate*}
This paradigm has several applications that range from
industrial control systems~\cite{sadeghi2015security} to small home-appliances~\cite{soliman2013smart}.

To enhance the security guarantees of IoT networks,
practitioners use Remote Attestation (RA) schemes
that allow a trusted entity (\ie \emph{Verifier}) to validate
the hardware and software integrity (\ie verifies) of 
a remote one (\ie \emph{Prover}) without the need of 
manual inspection.
The \emph{Verifier} sends a challenge to the \emph{Prover} and 
receives a measurement of its status as a report 
(\eg an application fingerprint). 
The \emph{Verifier} then compares the measurement 
with a database of \emph{proofs} previously saved to check 
the correctness of the \emph{Prover} status. 
This approach, often defined as \emph{single-Verifier RA}, is
well-established but assumes a \emph{Verifier} with powerful
capabilities and physically isolated from the network, thus protected against 
any threat (\eg a remote server in a controlled area).
The \emph{Prover}, instead, might be any device inside the network and can be 
tampered by a potential attacker. 
These assumptions, however, are incompatible in a scenario of 
independent IoT networks in which the devices cannot rely on an external \emph{Verifier} for the attestation process.
Moreover, an IoT network is usually composed by
devices that are equipped with different hardware and run 
different applications. 
A naive solution for removing a central \emph{Verifier} might 
be to keep a copy of all the \emph{proofs} inside each device. 
However, this is not practical due to the resource constraints 
imposed by IoT devices themselves.
Moreover, a dynamic network is composed by devices which 
continuously leave and enter the network itself, thus 
introducing additional \emph{proofs}-management challenges.

First solutions for autonomous networks were recently proposed
with~\cite{diat,us-aid,pasta}.
However, we found some limitations in these works.
In~\cite{diat}, the authors rely on heuristics and consider
only homogeneous devices, while in~\cite{us-aid,pasta}, they do 
not fully address scalability and still suffer from single
point-of-failure issues.
We provide more details in Section~\ref{sec:related-work}.


In this paper, we propose $\text{SAFE}^d$: the first concrete 
RA schema for autonomous networks of heterogeneous embedded
devices.
$\text{SAFE}^d$ introduces a new approach that enables any IoT 
device in a network to validate the integrity of another random device without the need for a central \emph{Verifier}.
%
To achieve this goal, we design new solutions that fits the IoT realm.
The main concept regards the decentralization of the \emph{Verifier}'s duties 
by making all the network active during the attestation process.
The distributed nature of our solution increases the effort required for an attack.
To implement our strategy, we built $\text{SAFE}^d$ upon a
Distributed Hash Tables (DHT)~\cite{zhang2013distributed} that 
allows us to manage data structures spread over several entities.
For our proof-of-concept, we opted for 
Chord~\cite{chord-2001,chord-2003,chord-correctness} 
as DHT implementation, however, $\text{SAFE}^d$ 
is agnostic from the type of DHT protocol chosen.
Chord protocol has not been thought to be resilient against 
compromised devices.
Therefore, we developed new solutions to improve DHTs 
security guarantees and to overcome their limitations.
First, we design a new key exchange protocol witch is based 
on Diffie Hellman~\cite{steiner1996diffie} 
and is integrated inside the Chord mechanisms.
It allows to issue a secure communication between any 
pair of devices inside the network without having to store cryptographic keys or using additional exchanged messages.
Second, we improve data resilience by using parallel 
DHT instances (called \emph{overlays}) that replicate 
the \emph{proofs} in different and random devices.
This avoids an attacker to infer the \emph{proofs} position 
inside the network.
Our schema requires every device to be equipped with 
small trusted anchors~\cite{10.1007/978-3-319-40385-4_6}, 
which contain and protect the algorithms required to $\text{SAFE}^d$.



We implemented $\text{SAFE}^d$ in the open-source
Raspberry Pi 3 platform. 
We chose this solution because its chip supports ARM 
TrustZone~\cite{winter2008trusted},
which is a standard trusted anchor largely used in other works~\cite{Eldefrawy:2017:HHD:3098243.3098261,10.1007/978-3-319-40385-4_6,Li:2015:ASO:2742647.2742676}.
Moreover, we performed a large-scale experimentation by simulating networks of 
$10$K virtual devices through Omnet++~\cite{omnetpp}.
To validate our approach, we conducted several attacks against both the 
platforms, encompassing software tampering, lost packets and corrupted 
devices. $\text{SAFE}^d$ recovered up to $99.9\%$ of lost 
\emph{proofs} and showed a logarithmic communication overhead 
and memory footprint.

$\text{SAFE}^d$ overcomes previous attestation solutions 
for network of IoT devices because its performances are not
affected by the number of devices connected and it completely
removes single point-of-failure by design.
Furthermore, solid experimental results are proposed 
to support our claims.
We believe that $\text{SAFE}^d$ will help developing more 
resilient networks of IoT devices and secure DHTs.

To sum up, $\text{SAFE}^d$ is a novel collaborative 
attestation for networks of heterogeneous IoT devices that
introduces the following contributions:
\begin{itemize}
	\item \textbf{No single point of failure}: the \emph{proofs}
	are randomly spread inside the network, thus increasing the
	difficulty for an attacker to corrupt the attestation
	process.
	\item \textbf{Self-protection}: the network can identify and react against corrupted devices.
	\item \textbf{Resilient network}: $\text{SAFE}^d$ can 
	recover its \emph{proofs} in case of lost data or attacks.
	\item \textbf{Scalability}: the protocol can manage a large 
	number of devices with minimal footprint.
\end{itemize}

The open-source proof-of-concept implementation of $\text{SAFE}^d$ for Raspberry Pi 3 will be available at the link~\footnote{We  are  willing  to  share the source code with the community upon acceptance or to provide it to the reviewers upon request via conference chairs}.

\section{Related Work}
\label{sec:related-work}


\subsection{Remote Attestation for IoT Networks}
\label{ssec:stor-remote-attestation}

The first attestation proposal for autonomous network 
was introduced by Abera et al. with DIAT~\cite{diat}. 
In this work, the authors assume a network of 
homogeneous devices (\eg a swarm of drones), which validate 
their own status without using a central \emph{Verifier}. 
However, they mainly focus on runtime RA (\ie they validate 
runtime device status) by using heuristics.
In addition, they require a network of homogeneous devices.
On the contrary, $\text{SAFE}^d$ is based on analytical results
and it can handle networks of heterogeneous devices in an autonomous fashion.

PASTA~\cite{pasta} is the first work that tries to spread 
the burden of verification across the entire network.
In PASTA, the \emph{Provers} periodically collaborate 
to generate the so-called \emph{tokens}.
Every \emph{token} attests the integrity of all the nodes 
that participated in its generation and contains a timestamp 
to allow absence detection of a particular node.
\emph{Tokens} are validated using an aggregated signature 
built on a Schnorr-based multisignature scheme.
However, we found some limitation to their approach since 
they require that each device maintains all the private keys 
of its neighbors, thus limiting the scalability. 
On the contrary, $\text{SAFE}^d$ enables any pair of devices 
to issue a trusted channel without any pre-shared information.

Another line of research attempts to measure the 
integrity of the network through a single challenge/response 
interaction~\cite{seda,sana,lisa}.
The common trait of these works is the logical organization 
of the devices using a spanning tree topology. 
An external \emph{Verifier} initiates the attestation process 
and collects a cumulative response from the entire network.
On the contrary, $\text{SAFE}^d$ faces a different scenario
and does not require a central \emph{Verifier} 
to perform an attestation.

Other works investigate cumulative RA schemes to detect physical 
attacks~\cite{darpa,scapi,us-aid,eapa}.
The intuition behind these solutions is that an adversary 
needs to remove a device from the network to perpetrate 
an attack, thus causing a temporary disconnection.
In particular, Ibrahim et al. proposed US-AID~\cite{us-aid} 
which combines continuous in-network attestation and
Proofs-of-non-Absence to detect both software tampering 
and device disconnections.
However, they require a reliable read-only clock (RROC) 
to achieve physical attacks detection.
Yan et al. improved the work of Ibrahim by introducing 
EAPA~\cite{eapa}, which performs RA physical attack resilient 
in a faster manner.
$\text{SAFE}^d$, instead, can detect physical attacks 
directly using communication timeouts by design. 

To sum up, $\text{SAFE}^d$ overcomes previous related RA schemes 
for mainly three reasons:
\begin{enumerate*}[label=(\roman*)]
	\item we fully remove any central trusted authority in the network,
	\item we efficiently spread the proofs among the nodes,
	\item we do not rely on synchronized clocks for absence detection.
\end{enumerate*}

\subsection{Distributed Hash Tables}
\label{ssec:dht-previous}
In general, all DHTs have been designed to 
decentralize information (\eg a file) and improve 
network performances and robustness.
Furthermore, they are thought to be deployed 
in large networks, such as the Internet.
However, these protocols do not consider security issues 
in their design.
In the last years, researchers investigated 
security limitations of
DHTs~\cite{kunzmann2006autonomically,sit2002security,fujii2009security,srivatsa2004vulnerabilities}.
These works aim at improving different aspects of 
DHT protocols, however, they differ from $\text{SAFE}^d$ 
for different reasons:
\begin{itemize}
	\item \textbf{Context:} they assume a large and 
	dynamic network such as the Internet, 
	while we focus on a more restricted physical area.
	Thus, we can rely on a better control 
	over the devices that compose the network.
	\item \textbf{Attacker model:} they consider 
	dishonest or churn nodes. On the contrary, $\text{SAFE}^d$
	assumes honest behavior guaranteed by trusted anchors.
	\item \textbf{Defense strategies:} they rely on 
	statistical and cryptographic schema 
	to improve trust in nodes~\cite{ahmat2019multipath}. 
	However, their approaches simply increase the effort required to a potential attacker without resolving the problem by design.
\end{itemize}
We are the first to tackle DHT security issues 
in the context of attestation protocols, 
which is more concrete and practical \wrt previous works.

Another crucial aspect of DHTs regards the privacy of 
the data stored inside of
it~\cite{o2004information,suvanto2005privacy}.
$\text{SAFE}^d$ enhances the privacy constraint 
by entirely encrypting the traffic and protecting 
sensitive memory locations inside every device.

To sum up, $\text{SAFE}^d$ improves DHT security guarantees 
by exploiting trusted computing for specific scenarios 
(\ie IoT).
Also, we believe that our solutions can be adopted 
to mitigate similar threats in more general scenarios.
\section{Background}
\label{sec:background}

\begin{figure*}[t]
	\centering
	\includegraphics[width=0.9\textwidth]{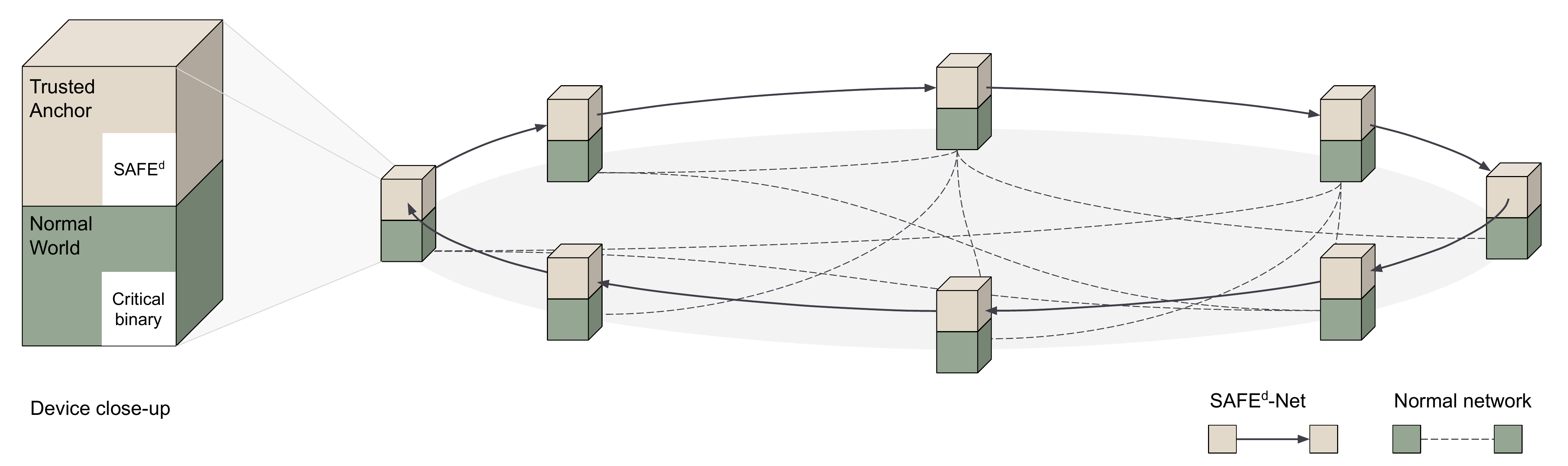}
	\caption{Network architecture (considering a single overlay).}
	\label{fig:network}
\end{figure*}


\subsection{Remote Attestation}
\label{ssec:remote-attestation}

Remote Attestation (RA) schemes refer to those protocols 
that allow verifying the integrity of a remote entity.
Usually, they involve two distinct roles: 
\emph{Verifier} and \emph{Prover}.
The \emph{Verifier} is considered trusted and is usually
physically protected from attacks (\eg a remote server).
Its duty is to verify the integrity of a \emph{Prover} 
that may be corrupted (\eg due to a malware).
RA schemes require a \emph{Verifier} to start the protocol 
by sending a challenge to the \emph{Prover}, which measures
some properties of its state (\eg compute a hash of a piece 
of software) and returns a report.
The \emph{Verifier} is now able to validate the \emph{Prover}
status according to the returned report by comparing 
its value with a database of correct measurements 
(called \emph{proofs}).

The classic approaches involve static measurements, 
such as software fingerprint or hardware
integrity~\cite{Eldefrawy:2019:ARA:3317549.3323403}.
More recently, researchers proposed a dynamic type of RA defined dynamic~\cite{abera2016c,diat,DBLP:journals/corr/abs-1807-08003}, 
which tries to attest run-time properties 
such as execution-paths.
$\text{SAFE}^d$ focuses on static RA. 
However, we discuss possible dynamic RA 
integration strategies in Section~\ref{sec:discussion}.

\subsection{Trusted Anchor}
\label{ssec:trusted-computing}

Modern RA schemes require devices to mount 
specialized hardware called trusted anchors.
These technologies allow to define protected 
memory regions and build Trusted Execution Environments (TEE).
A TEE provides useful functionalities like cryptographic algorithms and secure random number generators.
In this work, we opted for ARM
TrustZone~\cite{winter2008trusted} as trusted anchor 
due to its flexibility and its wide spread.
A device equipped with TEE organizes its memory 
in two main zones:
\emph{untrusted} and \emph{trusted}, respectively known 
in ARM TrustZone jargon as \emph{normal world} 
and \emph{secure world}.
The \emph{normal world}, as the name suggests, 
contains the general purpose software 
needed for running a classic operating system.
The \emph{secure world}, instead, contains the code 
strictly necessary for establishing a 
trusted execution inside the \emph{normal world}, 
\ie the \emph{secure world} checks 
the execution of the \emph{normal world}.

$\text{SAFE}^d$ implementation is placed inside 
the \emph{secure world} to protect its algorithms 
and critical variables (\eg cryptographic keys).
These technologies stand as the base of modern RA schema in IoT devices and classic IT infrastructure.

\subsection{Chord}
\label{ssec:chord}

Chord \cite{chord-2001,chord-2003} is a lookup protocol 
to establish a DHT and is specifically designed 
for large peer-to-peer networks.
This protocol allows to distribute hash tables 
(\ie key/value pairs) over multiple devices.
It uses consistent hashing for arranging the nodes 
in a circle and for distributing the keys among them. 
Each element (\eg nodes and keys) is identified by an
\textit{m-bit} number computed by a hash function.

Every piece of data is saved inside the first node 
whose identifier is greater than or equal to its identifier. 
As a consequence of this setting, a device will store all 
the piece of data whose ID is between its ID and that 
of the preceding device.
Each node is linked to its \emph{predecessor} 
and \emph{successor}, thus establishing a ring.
To improve resilience, a device also maintains a list 
of immediate successors, called \emph{successors list}.

Having the devices and data organized in a ring makes 
possible to implement a look up function as follows:
\begin{enumerate*}[label=(\roman*)]
	\item a device $A$ sends a request with the data ID 
	to its successor $B$,
	\item if $B$ contains the data (\ie $ID < B$), it is 
	returned to $A$, otherwise it forwards the request 
	to its successor. The step is repeated till finding 
	the ID.
\end{enumerate*}

The complexity of this solution is linear with the number 
of nodes placed inside the ring. 
To improve the performances, Chord introduces a 
\emph{finger table} that contains additional 
routing information. 
The \emph{finger table} has $m$ entries called 
\emph{fingers}: the i-th finger is a reference to the 
$2^{i-1}$ position ahead the current node.
As a result, the \emph{finger table} allows 
an average searching complexity of $O(\log_2 N)$, making 
the look up operation scalable with respect to the 
number of nodes $N$. 

Chord also provides procedures for adding new nodes 
to the ring and maintaining the order in case of failures.
We referred to~\cite{chord-correctness} for the 
implementation of a simpler yet correct version based 
on three distinct operations:
\begin{itemize}
	\item \textbf{Join}: an outside node contacts a member of the ring (defined as the \textit{entry}) to know 
	which is its successor. It then contacts its successor 
	to update the successor list.
	\item \textbf{Stabilize}: the node asks to its successor
	information about the predecessor. It adopts this
	predecessor as its new successor if it is actually 
	closer than the current successor in the ring order. 
	In both cases, the node sends a final notification 
	to the successor. The successors list is updated with 
	the information coming from the contacted nodes.
	\item \textbf{Rectify}: in case of a received 
	notification, the node checks if its current predecessor 
	is still alive and then adopts the notifying member 
	as new predecessor if it is closer than the current
	predecessor or if it has no live predecessor.
\end{itemize}
A node executes the \emph{join} procedure just when 
entering the network.
\emph{Stabilize} and \emph{rectify} procedures are 
instead periodically triggered during the protocol routine.

$\text{SAFE}^d$ builds on top of an enhanced version of Chord,
in which a joining node can save its attestation data 
inside the network and the routine operations take care of
re-distributing the information when new nodes join the ring. 
More importantly, $\text{SAFE}^d$ introduces redundancy of 
data by running several Chord instances at the same time, 
thus dealing with the loss of information caused by failures 
or attacks.
\section{Assumptions and Threat model}
\label{sec:threat-model}

%
%

\subsection{Device Architecture}
\label{ssec:device-architecture}

The devices considered in $\text{SAFE}^d$ are equipped with 
a trusted anchor (\ie ARM TrustZone), which 
is considered secure.
The \emph{secure world} is physically isolated from 
the rest of the system and its duties are twofold:
\begin{enumerate*}[label=(\roman*)]
	\item inspect and measure the device memory and
	\item communicate with the other trusted anchors 
	in the network.
\end{enumerate*}
The trusted anchor is used as a secure storage for 
all the variables needed by $\text{SAFE}^d$ and 
it is protected from an attacker by design.
The \emph{normal world} runs different applications and 
can be compromised.

\subsection{Network Context}
\label{ssec:network-context}

In this work, we assume networks of fully interconnected 
devices that range from few elements to $10$K devices.
Our main use case is for industry, however, we can 
deploy $\text{SAFE}^d$ to any type of autonomous 
system networks.
$\text{SAFE}^d$ can handle highly dynamic networks 
where nodes continuously enter and exit them.
However, we allow only known devices to join the network. 
This is reasonable since we consider geographically restricted networks (\eg factories or smart-homes).

\subsection{Threat Model}
\label{ssec:adverary-properties}

$\text{SAFE}^d$ faces attacks that target both the device 
and the network.

\paragraph*{\textbf{Device Attacks}}
The goal of the attacker is to load unauthorized binaries 
or inject malicious code inside the \emph{normal world} 
by using different strategies, \eg exploiting security flaws.
We consider the \emph{secure world} isolated from the
\emph{normal} one and therefore out of the attacker range.
We also consider compromised devices that hide their 
presence in the network and physical attacks.

\paragraph*{\textbf{Network Attacks}}
An attacker can manipulate network traffic by following 
classic Dolev-Yao model~\cite{dolev}. 
Thus she can eavesdrop, insert, modify, delete messages, 
perform a replay attack, or forge attestation messages.

\vspace{3mm}
In general, we do not consider denial-of-service (DoS), 
however, we evaluate the resilience of $\text{SAFE}^d$ 
in case of unavailable devices.
These assumptions are coherent with previous works~\cite{seda,sana,lisa,diat,us-aid,pasta}.
\section{$\text{SAFE}^d$}
\label{sec:safed}

\begin{figure*}[t]
	\centering
	\includegraphics[width=0.8\textwidth]{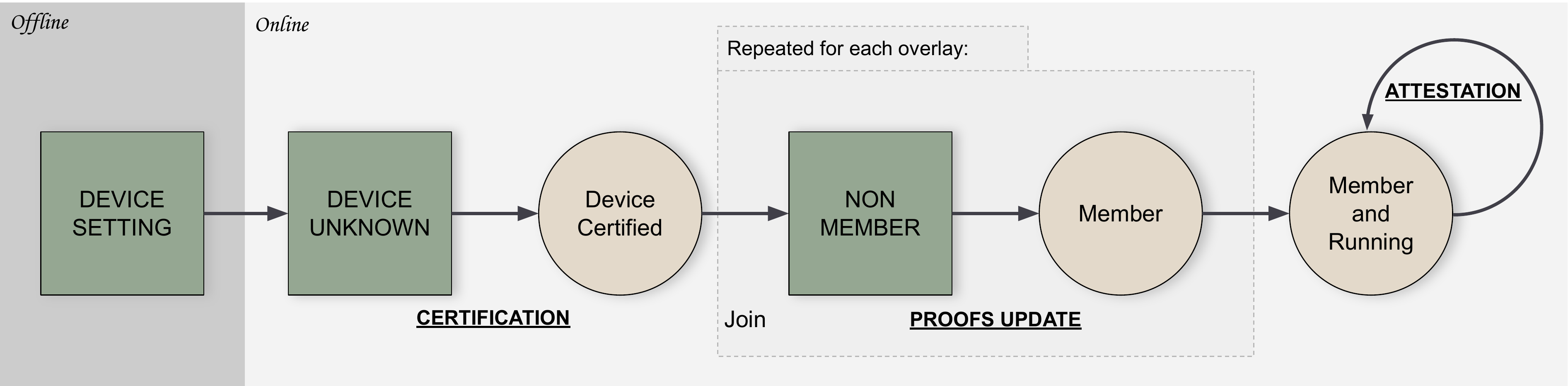}
	\caption{$\text{SAFE}^d$ finite-state-machine. Addition of states and 
	phases \wrt Chord are respectively shown with squares and bold underlined 
	text.}
	\label{fig:safed-steps}
\end{figure*}

$\text{SAFE}^d$ is an extension of a DHT that includes 
new mechanisms for dealing with the adversary 
described in Section~\ref{ssec:adverary-properties}.

In our schema, the devices are logically organized into two parallel networks (Figure~\ref{fig:network}). 
The first one is the normal network (dashed line), used by
ordinary applications to communicate with other devices.
The second one is \emph{$\text{SAFE}^d$-net} 
(solid-line arrows), built on top of the normal network 
and used by the trusted anchors to perform the schema 
routines.

\emph{$\text{SAFE}^d$-net} is composed by 
a fixed number of \emph{overlays} which is defined at 
network creation time.
Each \emph{overlay} is an independent Chord instance 
(Section~\ref{ssec:chord}) that contains all the devices 
as well as a copy for each \emph{proof}.
The purpose of the \emph{overlays} is twofold:
\begin{enumerate*}[label=(\roman*)]
	\item the number of \emph{overlays} identifies the redundancy (\ie $X$ independent \emph{proof} copies requires $X$ \emph{overlays}) and
	\item they helps keeping the \emph{proofs} distribution balanced.
\end{enumerate*}

The location of objects inside \emph{$\text{SAFE}^d$-net} 
is managed through two types of ID:
\begin{itemize}
	\item \textbf{Overlay ID (OID)}: identifies a device 
	inside an \emph{overlay}. Therefore, a device has 
	a different OID for each \emph{overlay}.
	\item \textbf{Unique ID (UID)}: identifies a 
	\emph{proof} of a device. It is the same for all 
	the \emph{overlays}.
\end{itemize}
Our design allows to randomly allocate the devices 
in each \emph{overlay}, and therefore, an adversary cannot predict the \emph{proofs} location.

We can explain the usage of these IDs by means of an example.
Assume $\text{SAFE}^d$ has two \emph{overlays}.
We assign three IDs to a device $D$, namely 
$D_{\text{OID}_1}$, $D_{\text{OID}_2}$, and $D_{\text{UID}}$.
The first two (\ie  $D_{\text{OID}_1}$ and $D_{\text{OID}_2}$)
identify the position of $D$ inside 
the two \emph{overlays} respectively. 
The last one (\ie $D_{\text{UID}}$) identifies the position
of its \emph{proof} in each overlay.
The OIDs are computed online when the device enters 
the network.
In Section~\ref{ssec:secure-device-comm}, we describe the
OID creation process that makes them random and applicable 
as public keys in a secure communication protocol.
The UID, instead, is computed offline and is used as 
the key for retrieving the \emph{proof} from the various
overlays during an attestation process.
We indicate UID and \emph{proof} as a key/value pair $(\text{UID},\emph{proof})$.
As a result, even if the UID is predictable, an external attacker cannot foresee which device contains the \emph{proof} linked with it.

\subsection{Protocol Overview}
\label{ssec:challenges}

In $\text{SAFE}^d$, we strengthen
the Chord protocol to achieve the following properties.
First, we desire that only authorized devices join the network.
Second, we need to spread multiple copies of the 
\emph{proofs} among the \emph{overlays}.
The entire protocol is represented as a finite-state-machine 
in Figure~\ref{fig:safed-steps}, where the shapes 
(\ie circle and squares) represent a device status, 
while the arrows represent the transaction from a status 
to the next one.
More precisely, the new status introduced by 
$\text{SAFE}^d$ are depicted as squares, while the new 
phases are labeled with bold underlined text.
In the following subsections, we describe the additions 
in detail:
\begin{itemize}
	\item \textbf{Device-setting} defines the initial 
	device configuration.
	\item \textbf{Device-unknown} and \textbf{Certification}
	allow only recognized devices to join the network 
	(Section~\ref{ssec:certification-phase}).
	\item \textbf{Non-member} and \textbf{Proofs update} 
	are used to enhance availability of data in our dynamic
	context (Section~\ref{ssec:membeship}).
	\item \textbf{Attestation} is used to monitor 
	the integrity of other devices inside the network
	(Section~\ref{ssec:attestation}).
\end{itemize}
As a whole, the protocol of $\text{SAFE}^d$ is composed by two distinct phases: \emph{Offline} and \emph{Online}.

During the \emph{Offline} phase, we boot the devices 
and set the following parameters inside the trusted anchor:
\begin{itemize}
	\item The pair $(\text{UID},\emph{proof})$, which will 
	be saved inside the network and later used for
	attestation/verification.
	\item A public/private key pair which are signed by 
	a certification authority (CA).
	\item The CA certificate for the key pair.
\end{itemize}

During the \emph{Online} phase, a device in \emph{Device-unknown} status connects to the network and 
starts the procedure to access \emph{$\text{SAFE}^d$-net}.
We introduced this phase because Chord does not handle 
authentication by default.
After the \emph{Certification} is done, a device enters the 
\emph{Device-certified} status and it can \emph{join} 
each \emph{overlay} asynchronously.
A device that does not pass the \emph{Certification} cannot physically communicate with other devices because it does 
not receive any OIDs from the entry point.
The following procedure is repeated for every \emph{overlay}.
During the \emph{join} phase, a device finds its 
successor around the ring by following standard Chord 
algorithms.
It then sends its pair $(\text{UID}, \emph{proof})$ to be 
stored.
After this task is completed, the device assumes a 
\emph{Non-member} status, which means that:
\begin{enumerate*}[label=(\roman*)]
	\item the device is aware of its position around the ring,
	\item it has inserted its own \emph{proof} inside 
	the \emph{overlay} and
	\item it has not received yet the \emph{proofs} it has 
	to store.
\end{enumerate*}
At this point, the device performs its first 
stabilize operation, making its successor aware of its 
presence inside the \emph{overlay}.
This triggers the \emph{Proofs update}, which consists 
in the successor sending a copy of the proofs that 
should be stored inside the new device.
A node maintains \emph{Non-member} status until it is
completely integrated inside the ring, \ie the preceding 
and following devices becomes aware of it. 
When this is the case, it has become officially part of the 
\emph{overlay} and it can switch to the \emph{Member} status.
This allows the device to perform the \emph{Rectify} 
operation as described in the original Chord protocol and permits to the successor to safely delete the proofs 
previously copied.
It is fundamental to maintain different \emph{Member} status 
for each \emph{overlay} because the \emph{Attestation} 
process involves all the \emph{overlays}. 
Therefore, we require a device to become \emph{Member} in all of them before being able to execute it, thus reaching
\emph{Member-and-Running} status.

\subsection{Certification Phase}
\label{ssec:certification-phase}
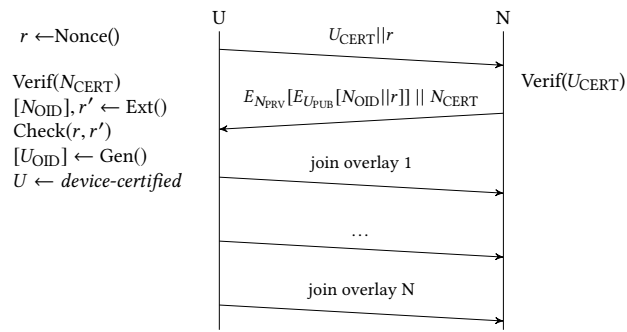
\begin{figure}[b]
	\centering
	\begin{adjustbox}{width=0.5\textwidth}
		\begin{tikzpicture}[node distance=4cm,auto,>=stealth']
		\node[right =0cm] (unknown) {U};
		\node[right =4cm of unknown] (entry) {N};
		\node[below of=unknown, node distance=5cm] (unknown_ground) {};
		\node[below of=entry, node distance=5cm] (entry_ground) {};
		\draw (unknown) -- (unknown_ground);
		\draw (entry) -- (entry_ground);
		\node[text width=2.8cm] at (-1.5,-0.3) {$r\gets$Nonce()};
		\draw[->] ($(unknown)!0.10!(unknown_ground)$) -- node[above,scale=0.9,midway,yshift=0.12cm]{$U_\text{CERT}||r$} ($(entry)!0.15!(entry_ground)$);
		\node[text width=2.3cm] at (6.1,-1) {Verif($U_\text{CERT}$)};
		\draw[->] ($(entry)!0.30!(entry_ground)$) -- node[above,scale=0.9,midway,yshift=0.12cm]{$E_{N_\text{PRV}}[E_{U_{\text{PUB}}}[N_\text{OID}||r]]\ ||\ N_\text{CERT}$} ($(unknown)!0.35!(unknown_ground)$);
		\node[text width=3cm] at (-1.5,-1.8) {$\text{Verif}(N_\text{CERT})$\\$[N_\text{OID}],r'\gets\text{Ext}()$\\$\text{Check}(r,r')$\\$[U_\text{OID}]\gets\text{Gen}()$\\$U\gets\emph{device-certified}$};
		\draw[->] ($(unknown)!0.50!(unknown_ground)$) -- node[above,scale=0.9,midway,yshift=0.12cm]{join~overlay~1} ($(entry)!0.55!(entry_ground)$);
		\draw[->] ($(unknown)!0.70!(unknown_ground)$) -- node[above,scale=0.9,midway,yshift=0.12cm]{\dots} ($(entry)!0.75!(entry_ground)$);
		\draw[->] ($(unknown)!0.90!(unknown_ground)$) -- node[above,scale=0.9,midway,yshift=0.12cm]{join~overlay~N} ($(entry)!0.95!(entry_ground)$);
		\end{tikzpicture}
	\end{adjustbox}
	\caption{Certification phase between an unknown device $U$ and the entry point $N$.}
	\label{fig:certification-phase}
\end{figure}
We desire that only authorized devices join  
\emph{$\text{SAFE}^d$-net} network because this ensures 
an honest execution of the \emph{$\text{SAFE}^d$-net} procedures (\ie due to trusted anchor isolation).
More precisely, an entry point $N$ can recognize the identity of a device $U$, which is in \emph{Device-unknown} status, 
by using a public key infrastructure (PKI), as described 
in Figure~\ref{fig:certification-phase}.
All the devices are initialized during the 
\emph{Offline phase} and receive a private key 
(\eg $N_\text{PRV}$) and the corresponding certificate 
(\eg $N_\text{CERT}$), signed by a certification 
authority (CA).
The procedure uses a generic asymmetric encryption schema denoted as $E$.

The protocol starts with $U$ that generates a nonce $r$ and sends it along with its certificate (\ie $U_\text{CERT}||r$) 
to $N$.
After $N$ correctly verifies the signature of $U_\text{CERT}$,
it encrypts all of its OIDs and the nonce $r$ 
(\ie $[N_\text{OID}||r]$) by first using the public key 
of $U$ (\ie $U_\text{PUB}$) and then its private key 
(\ie $N_\text{PRV}$).
Finally, $N$ sends them back to $U$ along with its 
certificate (\ie $N_\text{CERT}$).
At this point, $U$ performs the following operations:
\begin{enumerate*}[label=(\roman*)]
	\item verifies the certificate of $N$ (\ie $N_\text{CERT}$)
	using the CA public key,
	\item verifies the public key of $N$ using the 
	certificate and removes first encryption with it,
	\item extracts the OIDs of $N$ (\ie $[N_\text{OID}]$) 
	and nonce (\ie $r'$) using its own private key, 
	\item checks the nonce $r$ and $r'$ to avoid replay 
	attacks,
	\item generates its own OIDs (\ie $[U_\text{OID}]$), and
	\item sets its status to \emph{Device-certified}.
\end{enumerate*}
The double encryption guarantees two properties:
\begin{enumerate*}[label=(\roman*)]
	\item the public key of $U$ ensures that only $U$ 
	can decrypt the message and,
	\item the private key of $N$ ensures that the message has been sent by $N$.
\end{enumerate*}

From this point ahead, $U$ can communicate with the 
entry point by using the encryption schema described in Section~\ref{ssec:secure-device-comm}.
More precisely, $U$ \emph{joins} the \emph{overlays} 
as described in Chord.
A device that does not pass the \emph{Certification} 
phase cannot receive the OIDs of the entry point and, therefore, cannot communicate with the other devices.
To protect from the leakage of the private key during 
physical attacks, it is fundamental to implement a 
certificate revocation procedure. We discuss possible 
solutions in Section~\ref{sec:discussion}.

\subsection{Multiple Device Entrance}
\label{ssec:membeship}

$\text{SAFE}^d$ maintains all the \emph{proofs} available 
and consistent in all the \emph{overlays} in case of groups 
of devices that attempt joining the network simultaneously. 
To achieve this, we introduce the \emph{Non-member} status 
and the \emph{Proof update} task.

\begin{figure}[t]
	\centering
	\includegraphics[width=0.4\textwidth]{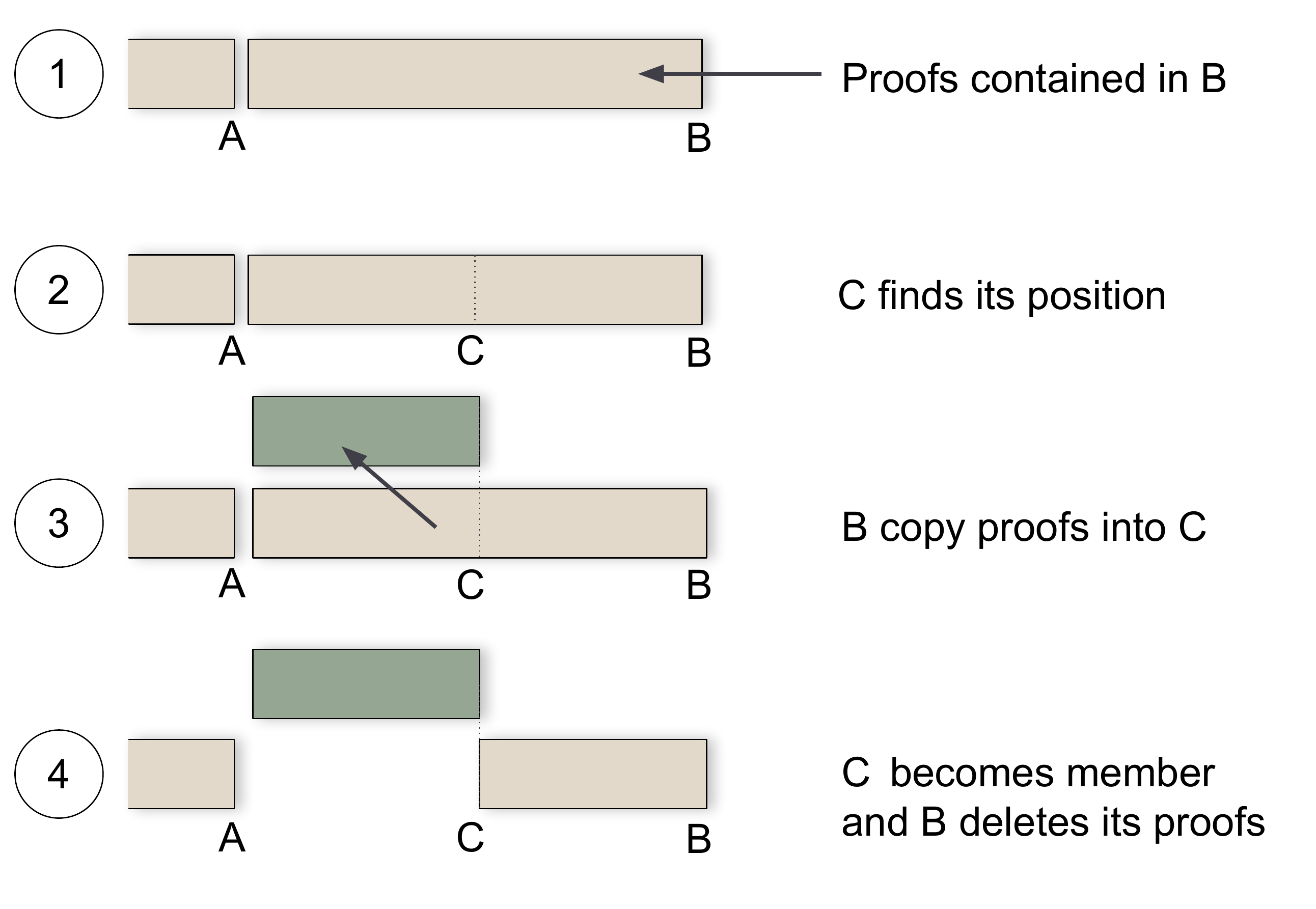}
	\caption{Main steps of a new device that enters the \emph{overlay}. 
		The device remains \emph{Non-member} until its neighbours in the overlay become aware of it, \ie it becomes \emph{Member} of the \emph{overlay}.}
	\label{fig:membership}
\end{figure}

Figure~\ref{fig:membership} shows the main four steps 
of \emph{Proof update}, that begins when a new device 
$C$ enters in an \emph{overlay}.
The rectangle before the node letter is a representation of 
the \emph{proofs} stored inside of it.
In step~1, we assume having two devices $A$ and $B$ 
correctly distributed around the \emph{overlay} 
(\ie $A_{\text{OID}} < B_{\text{OID}}$).
In step~2, $C$ has just \emph{joined} and it has found 
its position between $A$ and $B$
(\ie $A_{\text{OID}} < C_{\text{OID}} < B_{\text{OID}}$).
In this step, $C$ is in \emph{Non-member} status and it can 
only perform \emph{stabilize}.
This allows us to handle the entrance of multiple 
devices simultaneously and will be described later.
In step~3, $B$ copies into $C$ the relative \emph{proofs} 
while keeping a temporary copy in $B$ itself.
Keeping a copy of the \emph{proofs} into $B$ enables the 
other devices to find the $C$ \emph{proofs} even though 
$C$ has not entered the ring yet.
In step~4, $A$ performs rectify and inserts $C$ as 
its successor.
$C$ is formally part of the \emph{overlay} and consequently 
can shift to \emph{Member} status, while $B$ can delete 
its leftover \emph{proofs}.
After step~4, $C$ starts performing \emph{rectify}.
When $C$ becomes \emph{Member} in all the \emph{overlays}, 
it reaches \emph{Member-and-running} status and starts performing/receiving attestations.

This approach allows us to handle group of devices that
enter simultaneously.
For instance, a new device $D$ may attempt entering 
while $C$ is still a \emph{Non-member}. 
Here, we distinguish two cases:
\begin{enumerate*}[label=(\roman*)]
	\item $D$ is located between $A$ and $C$ 
	(\ie $A_{\text{OID}} < D_{\text{OID}} < C_{\text{OID}}$) 
	and sets $C$ as successor;
	\item $D$ is located between $C$ and $B$ 
	(\ie $C_{\text{OID}} < D_{\text{OID}} < B_{\text{OID}}$) 
	and sets $B$ as successor.
\end{enumerate*}
In both cases, $D$ is kept as \emph{Non-member} until 
its successor becomes \emph{Member} as well.
The difference is just in the order in which the devices 
become \emph{Member}.
In case (i), $D$ receives \emph{Member} status from $C$, therefore, the entrance order is $C$, then $D$.
In case (ii), $D$ receives \emph{Member} status from $B$, 
while $C$ changes its successor to $D$ after doing a
\emph{stabilize} to $B$.
Generally, \emph{Member} status is assigned only by other 
\emph{Member} (or \emph{Member-and-running}) devices, 
which are considered stable.
We keep \emph{rectify} disabled while a device is 
\emph{Non-member} to avoid the formation of \emph{chains} 
of devices that would cause some of them to have outdated
list of proofs.

\subsection{Attestation Protocol}
\label{ssec:attestation}

$\text{SAFE}^d$ attestation process is an extension of 
classic RA (Section~\ref{ssec:remote-attestation}).
The main differences are essentially two:
\begin{enumerate*}[label=(\roman*)]
	\item all the devices can assume the role of either 
	\emph{Prover} or \emph{Verifier}, and
	\item the \emph{proofs} are spread among all devices.
\end{enumerate*}
The entire attestation process is implemented in a 
dedicated phase, which is executed whenever a device inside 
the network needs to verify the integrity of another device
before starting a communication.

The attestation is composed by the following steps:
\begin{itemize}
	\item \textbf{Verifier initialization}: when the 
	attestation is triggered, the device enters in 
	\emph{Verifier} mode.
	\item \textbf{Attestation request}: the \emph{Verifier} challenges the 
	\emph{Prover}.
	\item \textbf{Retrieve report}: the \emph{Prover} 
	measures itself and returns a report defined as
	$(\text{UID}, \text{HASH})$ to the \emph{Verifier}.
	The UID is used to retrieve the \emph{proofs} from
	the \emph{overlays}, while the HASH is the current
	self-measure of the \emph{Prover}, which can be corrupted.
	\item \textbf{Retrieve proofs}: the \emph{Verifier} 
	queries all the \emph{overlays} to retrieve the \emph{proofs} using the received UID. The \emph{proofs} will be used to check the report validity.
	\item \textbf{Voting}: we use a First Past The Post (FTPT) 
	voting schema~\cite{fptp} to decide the healthy status of 
	the \emph{Prover}. In case of compromised status, 
	the \emph{Verifier} reacts in a proper way 
	(see Attack Reaction below).
	\item \textbf{Recovery}: in case of missing 
	\emph{proofs}, the \emph{Verifier} uses the retrieved 
	data to recover the information where it has been lost. 
	In case no \emph{overlay} returned the \emph{proof}, the 
	\emph{Verifier} launches a warning of possible infection for that particular device. Finally, the device exits from the \emph{Verifier} mode.
\end{itemize}

\paragraph*{\textbf{Voting}}
FTPT is a plurality voting system that elects the most 
voted choice as the winner.
During the voting phase, the \emph{Verifier}
considers the \emph{proofs} collected from the various 
\emph{overlays} as a preference vote, \ie which state each 
overlay thinks is the correct one.
At first, the \emph{Verifier} evaluates all the missing 
\emph{proofs} as blank votes and it does not consider 
them in the counting.
Then, the \emph{Verifier} chooses the correct \emph{proof} by
picking the one that was returned the highest number of times.
The elected \emph{proof} is then used to verify the 
\emph{report} sent by the \emph{Prover}.
The devices whose vote disagreed with the elected 
\emph{proof} are considered infected as well.

Despite its simplicity, FTPT shows in our case resilience 
against manipulation.
An attacker cannot foresee the location of the \emph{proofs} 
in the network.
As a result, the voting is robust till at least $50\%$ 
devices are healthy.
Due to the design of $\text{SAFE}^d$, we can easily implement new type of voting schema~\cite{tbyzantine}.



\paragraph*{\textbf{Attack Reaction}} The actual attack reaction strategy strictly depends by the network pursue. 
In our prototype, we isolate the corrupted devices.
In other scenarios, for instance, we can implement an
hard-reset of the device.
This can be useful for malware such as Mirai~\cite{mirai}.
It is also possible to save the attestation results in the 
\emph{overlays} for future manual inspections.

\subsection{Secure Device Communication}

\label{ssec:secure-device-comm}
$\text{SAFE}^d$ allows two devices to issue a 
secure communication channel by introducing a novel 
protocol that allows to share a symmetric key $K$ 
with zero-message exchanged.
Our approach overcomes the scalability limitations of previous 
ones~\cite{us-aid,pasta} that either requires
a device to store every key needed for the communication or 
to execute a key exchange protocol to establish 
a secure channel.
The protocol is based on Diffie 
Hellman~\cite{steiner1996diffie} and exploits 
Chord properties.
The main idea is that each OID represents the public key 
of a device that joined an \emph{overlay}.
In Section~\ref{ssec:certification-phase}, we discuss 
our mechanism to allow only authorized devices 
to enter the network.

In our protocol, we assume that all the devices 
share two secure prime number $g$ and $N$.
During the join phase, a device randomly computes 
an OID as follows:
\begin{equation*}
	\begin{split}
		X &= \text{rand}()~(\text{modulo}~N),\\
		\text{OID} &= (g^X)~(\text{modulo}~N).
	\end{split}
\end{equation*}
At first, a device randomly computes a number $X$ (modulo $N$), which is kept secret within the trusted anchor.
Then, it generates the OID by computing the exponentiation 
of $X$ over $g$ (modulo $N$).
These two operations are repeated for each \emph{overlay}.
For the sake of simplicity, we continue the 
description considering a single \emph{overlay}, however, 
it is possible to easily extend the approach to
any number of \emph{overlays}.
We indicate the pair $X$, OID for a device $D$ as follows:
\begin{equation*}
	\begin{split}
		(D_{X}, D_{\text{OID}}).
	\end{split}
\end{equation*}
Two devices, namely $A$ and $B$, that know the respective 
OIDs can compute a shared symmetric key $K_{AB}$ as follows:
\begin{equation*}
	\begin{split}
		K_{AB} &= (A_{\text{OID}})^{B_X} = {(g^{A_X})}^{B_X} \\
		K_{AB} &= (g^{B_X})^{A_X} = (B_{\text{OID}})^{A_X}~(\text{modulo}~N).
	\end{split}
\end{equation*}
The key $K_{AB}$ can now be used in a symmetric encryption schema $E$.

The design of Chord assures that each device knows 
the OIDs of its successors 
(\eg finger list and successor list).
Therefore, if $B$ is a device following $A$, 
$B$ cannot compute $K_{AB}$ because it does not have 
knowledge of $A_\text{OID}$.
To overcome this problem, we need to send $A_\text{OID}$ 
to $B$ avoiding unauthorized entities to read the OID. 
We achieve this by encrypting $A_{\text{OID}}$ 
such that only authorized devices can read it. 
The whole message structure is shown in the following:
\begin{equation*}
	\begin{split}
		E_{B_\text{OID}}[S_\text{OID}||A_\text{OID}]\  ||\  E_{K_{AB}}[M]\  
		||\  O,
	\end{split}
\end{equation*}
which comprises three parts:
\begin{itemize}
	\item $E_{B_\text{OID}}[S_\text{OID}||A_\text{OID}]$ 
	is the header and is encrypted by using a symmetric 
	encryption schema $E$ and $B_\text{OID}$ as a
	key~\footnote{Since the key space of $E$ is generally 
	smaller than the size of $B_\text{OID}$, we use a hash 
	function $H$ to adjust the size,  
	\ie $E_{H(B_\text{OID})}[.]$}.
	This allows only the devices that are already 
	participating the overlay (\ie $B$) to read the content.
	The header contains two OIDs, called \emph{source} 
	and \emph{sender}. 
	The first identifies the device which originally sent 
	the message (\ie $S_\text{OID}$),
	while the second identifies the device which is 
	currently forwarding the message to $B$ 
	(\ie $A_\text{OID}$). Keeping the \emph{source} OID 
	permits a fast reply; this will be described 
	later through an example.
	\item $E_{K_{AB}}[M]$ is the message body. It contains 
	the message $M$ to deliver and is encrypted with 
	the symmetric schema $E$ and $K_{AB}$ as a key.
	\item $O$ indicates the \emph{overlay} to which the message is meant to and is sent as plain text.
\end{itemize}
This structure enables $B$ to decrypt an incoming message 
as follow:
\begin{itemize}
	\item reads $O$ and identify from which \emph{overlay} 
	is coming.
	\item uses its corresponding OID (\ie $B_\text{OID}$) 
	to decrypt the header and to retrieve the sender OID 
	(\eg $A_\text{OID}$).
	\item computes $K_{AB}$ and decrypts $M$.
\end{itemize}

A packet structured in such way has three interesting properties:
\begin{enumerate*}[label=(\roman*)]
	\item besides the overlay $O$, no information is shipped 
	as a plain text, thus only devices inside
	\emph{$\text{SAFE}^d$-net} (also called \emph{Member}) 
	can read the headers (see 
	Section~\ref{ssec:certification-phase});
	\item only the intended recipient can successfully 
	decrypt the message, thus any attempt to manipulate 
	or reroute the message will generate an error;
	\item a symmetric schema is less expensive than 
	an asymmetric one, thus more suitable for 
	low-power devices.
\end{enumerate*}

\begin{figure}[t]
	\centering
	\includegraphics[width=0.3\textwidth]{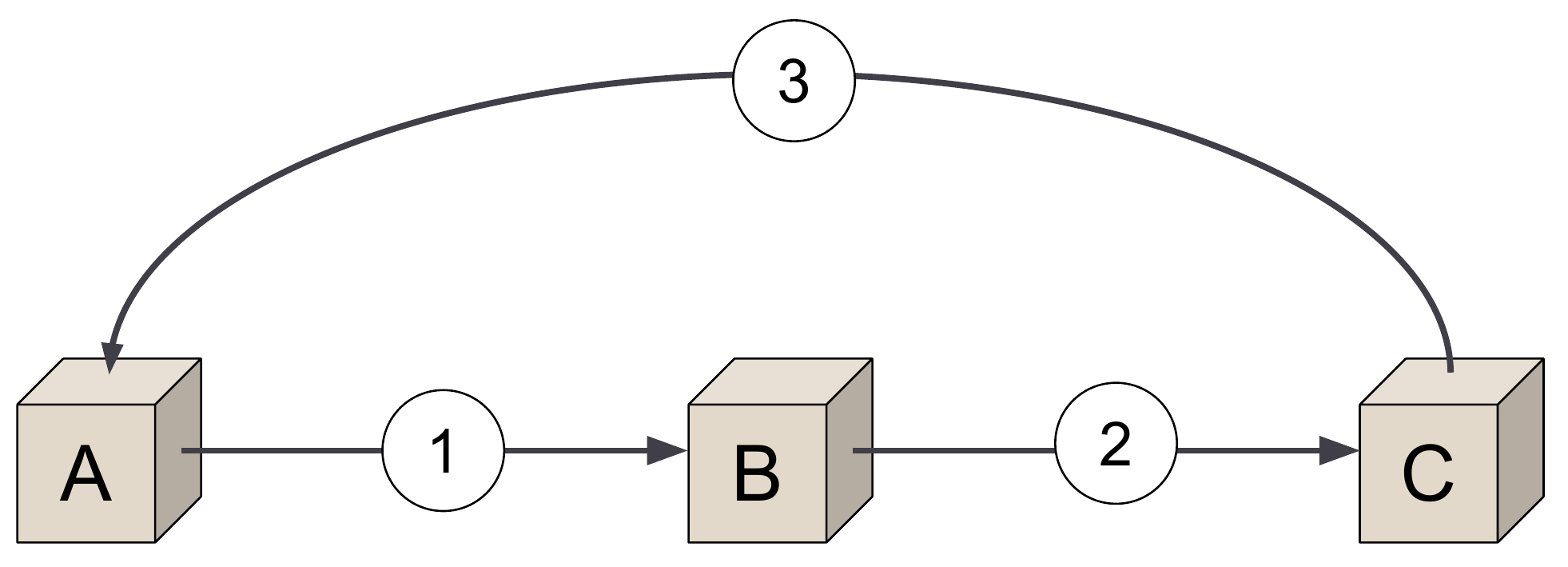}
	\caption{$\text{SAFE}^d$ point-to-point communication protocol.}
	\label{fig:ptp-protocol}
\end{figure}

Figure~\ref{fig:ptp-protocol} shows a complete example of two 
devices that communicate.
In this case, a device $A$ wants to send a message $M$ 
to device $C$, but $C$ is not directly reachable by $A$. 
Therefore, $A$ must pass through the ring.
At the beginning, $A$ only knows the OID of $B$, because it is 
its successor. Therefore, $A$ asks $B$ to deliver $M$ to $C$ 
by creating packet $(1)$ as follows:
\begin{equation*}
	\begin{split}
		E_{B_\text{OID}}[A_\text{OID}||A_\text{OID}]\  ||\  E_{K_{AB}}[M]\  
		||\  O.
	\end{split}
\end{equation*}
In this packet, \emph{source} and \emph{sender} OID 
coincide with $A_\text{OID}$.
$B$ decrypts the header with its OID, calculates $K_{AB}$ 
using the \emph{sender} OID and obtains the message $M$. 
Since $B$ knows $C$, $B$ relays the message crafted as follows:
\begin{equation*}
	\begin{split}
		E_{C_\text{OID}}[A_\text{OID}||B_\text{OID}]\  ||\  E_{K_{BC}}[M]\  
		||\  O.
	\end{split}
\end{equation*}
$C$ follows similar steps to retrieve message $M$
and serves the request.
At this point, $C$ replies to $A$ by using 
the \emph{source} OID (\ie $A_\text{OID}$) and 
crafting message $(3)$ as follows:
\begin{equation*}
	\begin{split}
		E_{A_\text{OID}}[A_\text{OID}||C_\text{OID}]\  ||\  E_{K_{CA}}[M]\  
		||\  O.
	\end{split}
\end{equation*}
Finally, $A$ receives the response from $C$.

This approach brings three advantages: 
\begin{enumerate*}[label=(\roman*)]
	\item we avoid spoofing attacks because the sender 
	is automatically verified
	(unless the attacker stoles its secret $X$),
	\item we can build a symmetric key without using 
	extra messages, and
	\item a compromised device cannot choose its 
	OIDs arbitrarily unless it resolves the 
	discrete logarithm problem.
\end{enumerate*}
We also mitigate reply attacks by using nonces~\cite{zhen2003preventing}.
\section{Implementation}
\label{sec:implementation}

\begin{figure}[t]
	\centering
	\includegraphics[width=0.45\textwidth]{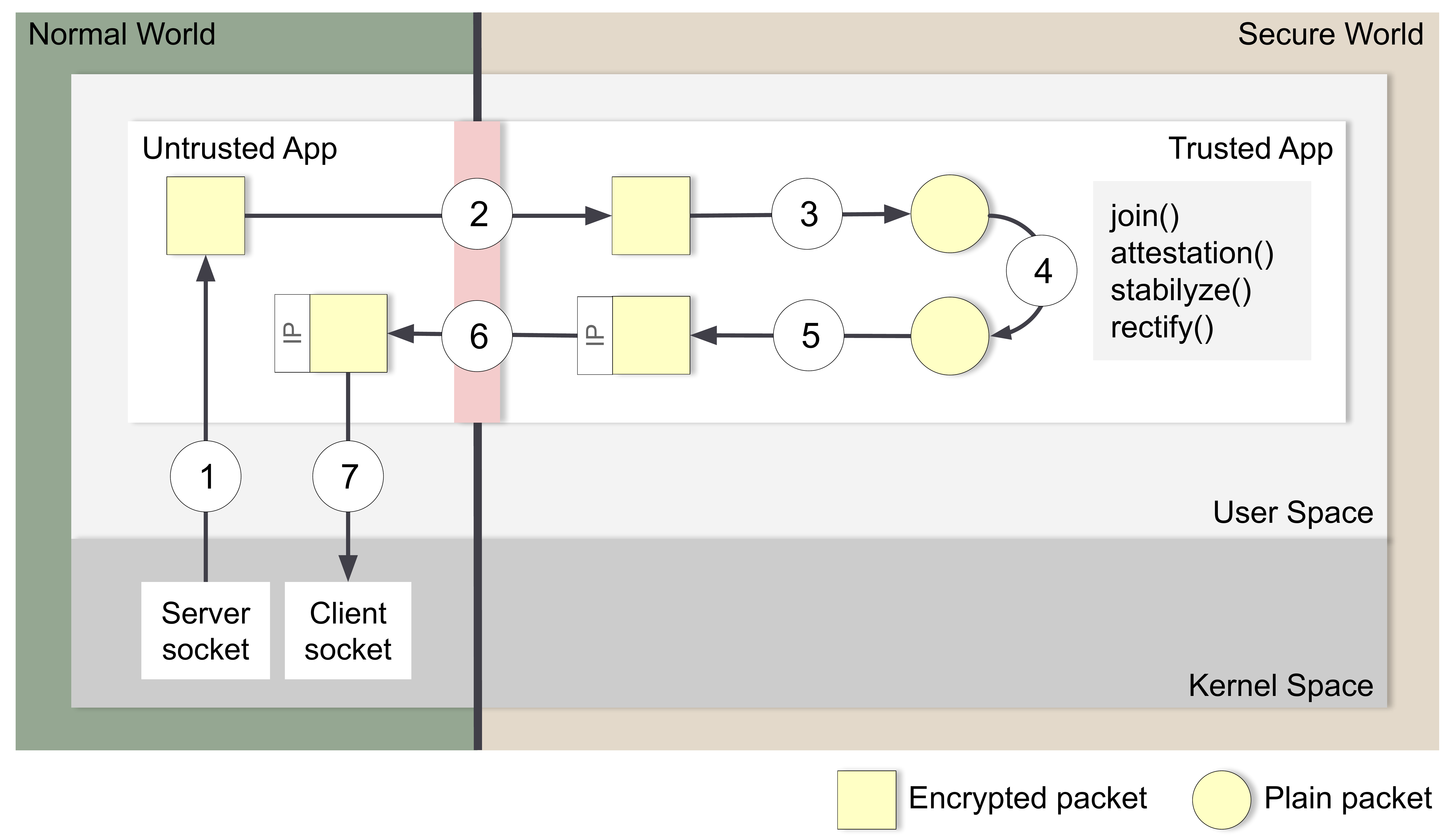}
	\caption{Execution pipeline.}
	\label{fig:architecture}
\end{figure}

Figure~\ref{fig:architecture} shows the architecture adopted 
for the platform Raspberry Pi 3. We developed our prototype 
on top of OP-TEE~\cite{optee}~\footnote{We used the commit 
\texttt{\seqsplit{f5172a4aa993f644d0edb3a64a49938fd2e6f906}} 
and \texttt{\seqsplit{28eea17f4dba5bbf7848926eb031ba660e8856f0}}
of the official repository for the OP-TEE OS and client
respectively.} and wrote it using C language.
Since we designed $\text{SAFE}^d$ to exploit ARM TrustZone
features (Section~\ref{ssec:trusted-computing}), 
we split $\text{SAFE}^d$ into two components: 
\emph{untrusted} and \emph{trusted application}.
The \emph{untrusted application} interacts with the 
peripherals, while the \emph{trusted application} 
contains $\text{SAFE}^d$ code and the private information.
The network communication is implemented through a TCP client
and server socket in the \emph{normal world}~\footnote{We could 
have implemented a socket in the secure world as well, however, 
not all trusted anchor platforms support this feature so we 
opted for a more flexible solution.}.
$\text{SAFE}^d$ workflow is composed by a number of 
independent steps, which are depicted in
Figure~\ref{fig:architecture}.
In the beginning, $\text{SAFE}^d$ waits for an incoming 
packet from the server socket (step~$1$). 
Once a packet arrives, it is sent into the 
\emph{trusted application} (step~$2$).
At this point, the packet is decrypted (step~$3$) and 
processed (step~$4$). After the response is created, it is
encrypted (step~$5$) and written into the 
\emph{untrusted application} along with the destination IP
(step~$6$), which must be in plain text for correct routing.
Finally, the packet is shipped by the client socket (step~$7$).
The packets are built in such a way that the 
\emph{untrusted application} knows only the destination IP,
while the content is always encrypted as described in
Section~\ref{ssec:secure-device-comm}.

An attacker that alters the plain-text IP would simply lead 
to a trashed or lost message because the only device capable 
of decrypting it is the intended recipient (see
Section~\ref{ssec:secure-device-comm}). Moreover, blocking 
the message would cause the original sender to raise a 
warning for a timeout in its communications, thus exposing 
the attack.

Our prototype requires around $46$KLoC for the untrusted
application and around $49$KLoC for the trusted application.
We used AES-CBC~\cite{thakur2011aes} for symmetric encryption 
with keys $32$B long, while we used
RSA~\cite{barrett1986implementing} for the asymmetric keys 
in the \emph{certification phase}, with $1218$B for the 
private key and $294$B for the public one.

\paragraph*{\textbf{Measurement Generation}}
In our proof-of-concept, $\text{SAFE}^d$ protects 
the integrity of critical pieces of software inside 
the \emph{normal world} by using a shared memory.
However, it is possible to extend $\text{SAFE}^d$ 
to measure other device properties, such as 
hardware configuration.
The location to protect is identified at the boot phase.
For the sake of simplicity, our proof-of-concept 
can monitor memory regions that reside in the same process 
of $\text{SAFE}^d$. 
It is still possible to extend $\text{SAFE}^d$ 
to read arbitrary physical addresses
and protect the integrity of different parts of the system~\cite{williams2015inspecting}.

\paragraph*{\textbf{Omnet++ simulation}}
We performed a large scale performance analysis using 
Omnet++~\cite{omnetpp,omnetppw} as support. We implemented 
our protocol at the application level and used time delays to
simulate cryptographic operations and propagation time.
Based on the Raspberry Pi's measurements, we set a delay of $\text{10}$ ms for 
the decryption/encryption of the messages.
Moreover, we set the communication rate at $\text{250}$ Kbps based on the defined data-rate of ZigBee, a widely used
communication protocol for networks of IoT devices.

\section{Evaluation}
\label{sec:evaluation}


\begin{figure*}
    \centering
     \begin{subfigure}[t]{0.48\textwidth}
        \centering
        \includegraphics[width=\textwidth]{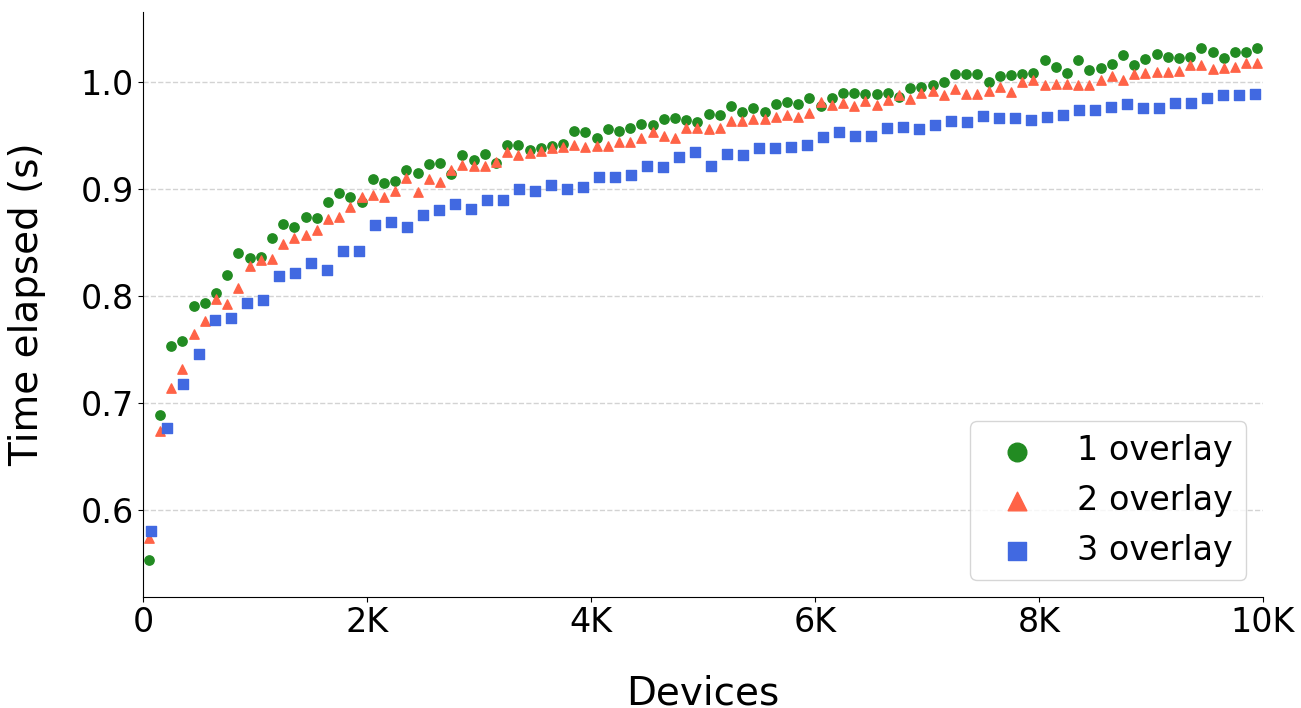}
		\caption{Time elapsed for performing a complete attestation.}
		\label{fig:messages}
    \end{subfigure}
    \hfill
	\begin{subfigure}[t]{0.48\textwidth}
		\centering
		\includegraphics[width=\textwidth]{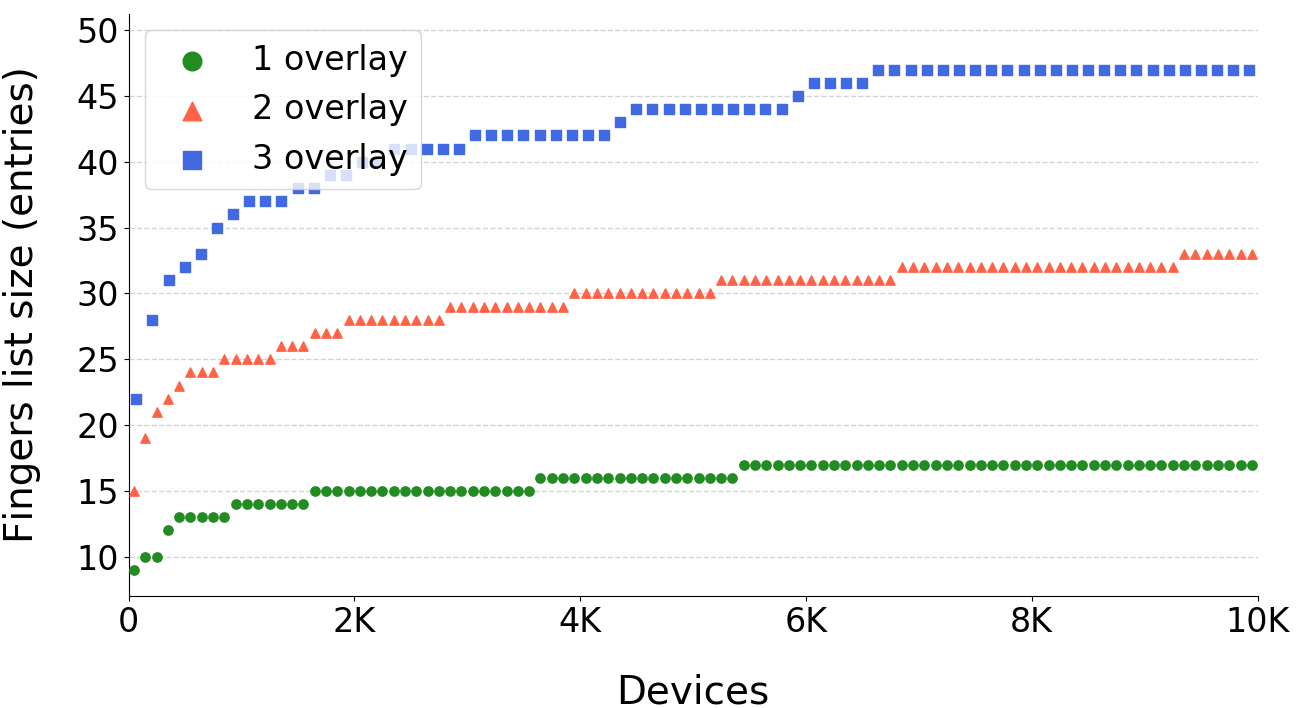}
		\caption{Maximum finger table size, 
		considering all the overlays.}
		\label{fig:fingertablesize}
	\end{subfigure}
    \hfill
    \vspace{1cm}
	\begin{subfigure}[t]{0.48\textwidth}
		\centering
		\includegraphics[width=\textwidth]{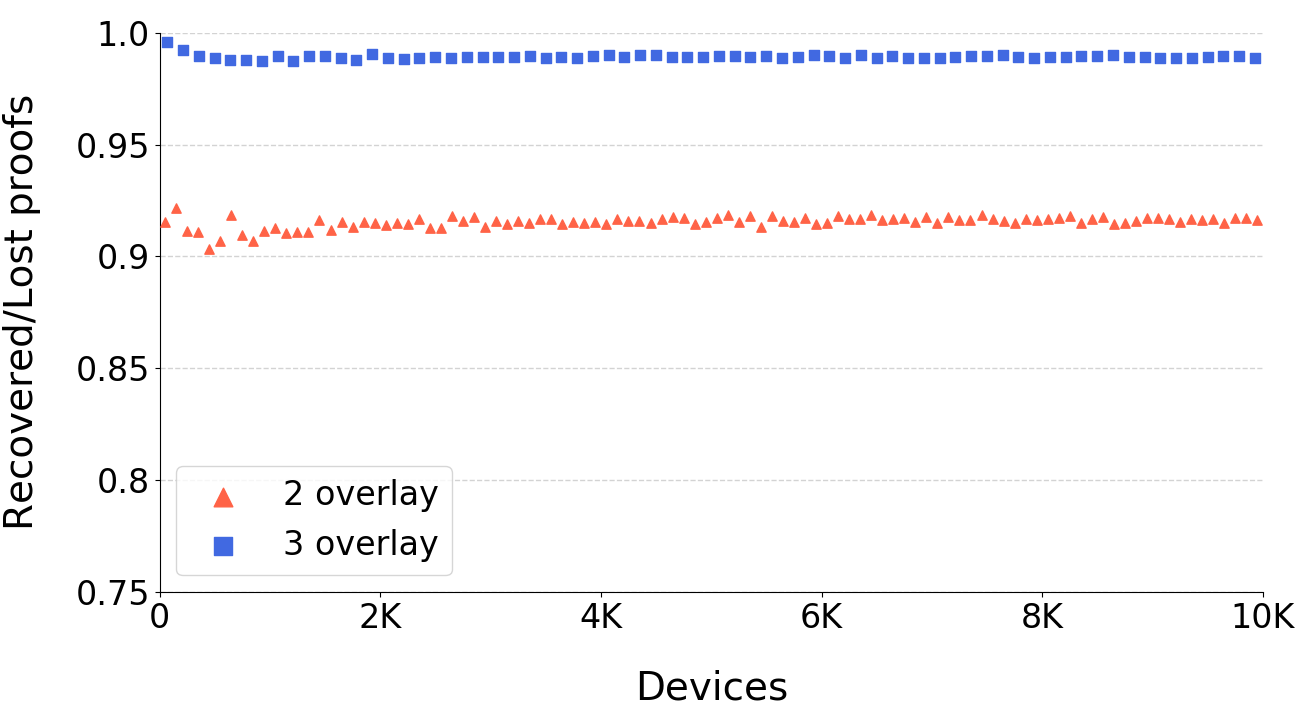}
		\caption{Resilience to lost proofs,
		measured as the recovered proofs over the
		lost ones with a drop-rate of 20\%.}
		\label{fig:resilience}
	\end{subfigure}
	\hfill
	\begin{subfigure}[t]{0.48\textwidth}
		\centering
		\includegraphics[width=\textwidth]{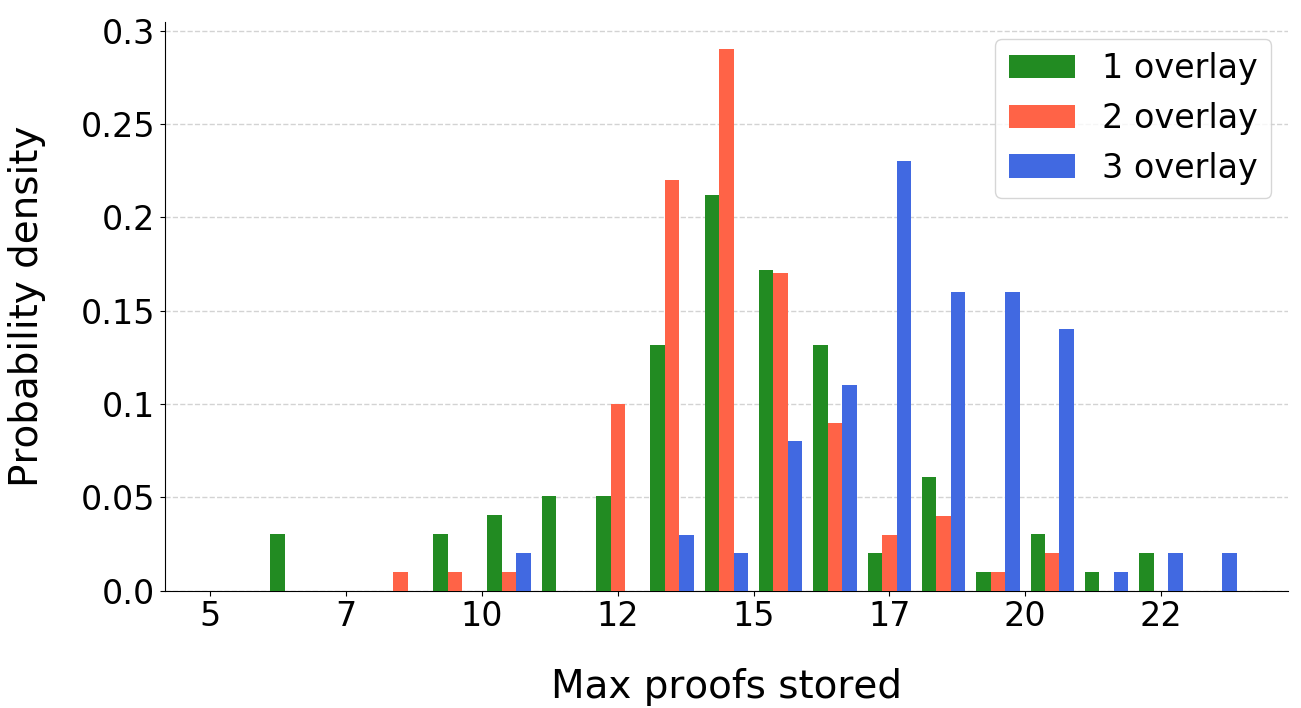}
		\caption{Probability density of the
		maximum number of proofs stored by a single device in a network of $10$K devices.}
		\label{fig:random}
	\end{subfigure}
	\caption{Network overhead, finger table size, resilience measurements and maximum 
	\emph{proofs} overhead  of $\text{SAFE}^d$ prototype.}
	\label{fig:three graphs}
\end{figure*}

In this section, we evaluate different metrics 
of $\text{SAFE}^d$ by using two network settings:
\begin{itemize}
	\item \textbf{Raspberry Pi}: we mounted a small network of Raspberry~Pi~3 
	composed by $4$ devices. This setting was used to test the
	efficacy of $\text{SAFE}^d$ on real devices and to collect realistic 
	parameters for the simulation.
	\item \textbf{Omnet++}: we used Omnet++ to simulate 
	a network of IoT devices that contained up to 
	$10$K entities and with different number of overlays 
	(from $1$ to $3$).
	We used this setting to evaluate the performances of
	$\text{SAFE}^d$ in the presence of thousands of devices 
	and with a different number of \emph{overlays}.
\end{itemize}

\subsection{Network Analysis}
\label{ssec:performance}

We measure the size of the message used by $\text{SAFE}^d$, 
the elaboration time, and the time elapsed to perform a complete attestation.

\paragraph*{\textbf{Messages Size.}} 
$\text{SAFE}^d$ requires two types of message:
\begin{itemize}
	\item \textbf{Certification messages}: they are used 
	only during the certification phase
	(Section~\ref{ssec:certification-phase}) and 
	require $256 \times o$ bytes, where $o$ is the number 
	of \emph{overlays} (\eg $1$KB for $4$ \emph{overlays}).
	These messages are more expensive in terms of size but 
	they are only used in the initial part.
	\item \textbf{Routine messages}: the other messages
	exchanged in our prototype have a fixed size of 
	$384$ bytes plus $68 \times s$ bytes, where $s$ is 
	the successor list size (\eg $520$B with $2$ successors).
	They compose the vast majority of the
	network communication.
	Furthermore, $s$ is a fixed parameter of the network, thus
	the message size remains constant throughout the execution.
\end{itemize}

\paragraph*{\textbf{Message Time Elapsed}}
We measure the time required to process a single message 
in a Raspberry Pi.
We did not consider the \emph{Certification} messages 
phase because they are used only during the initial part 
of the protocol.
As a result, any message required on average $9.1$ms (with 
a standard deviation of
$6.3$) to be processed without using cryptography,
while around the double, $18$ms (with a standard deviation 
of $9.2$), using the cryptography described in
Section~\ref{ssec:secure-device-comm}.
As already observed in~\cite{us-aid,pasta}, cryptography 
is the predominant part during the protocol execution.

\paragraph*{\textbf{Attestation Performances}}
Figure~\ref{fig:messages} shows the average time elapsed
to perform a complete attestation (y-axis) over the number
of devices present inside the network (x-axis).
A complete attestation involves sending the challenge,
collecting the report and querying all the \emph{overlays} 
to retrieve the \emph{proofs}.
The graph shows a logarithmic growth with the number 
of devices, while the number of \emph{overlay} does not affect
the overall performances because the packets are processed
by each device independently.
Considering $10$K devices and $3$ \emph{overlays}, 
an entire attestation process requires around
one second to be completed.

\subsection{Memory Footprint}
\label{ssec:memory-footprint}
Each device uses $64$, $1218$, $294$ and $256$ bytes 
respectively for the UID, the private RSA key, the public RSA 
key and the certificate.
For each overlay, it uses $32$ bytes for the secret, $68$ bytes
for its own OID and its predecessor OID, $68$ bytes for each 
entry in the successor list, $68$ bytes for the entries in the 
finger list and $128$ bytes for each element inside the proof 
storage.
The overall memory usage $M$, expressed in bytes, 
can be computed as follows:
\begin{equation*}
\begin{split}
M = 1832+[168+68\times{s}+68\times{f}+128\times{p}]\times{o},
\end{split}
\end{equation*}
where $s$ is the size of the successor list, $f$ is the size 
of the finger table, $p$ is the number of \emph{proofs} 
and $o$ the number of \emph{overlays}. 
Variables $s$ and $o$ are parameters that remain 
constant during the protocol execution.
In the following paragraphs, we show that the 
finger list size $f$ has an upper bound of $\log(n)$, 
with $n$ the total number of devices inside the network,
while the number of proofs $p$ is small \wrt the network size.

\paragraph*{\textbf{Finger Table Size}}
Figure~\ref{fig:fingertablesize} shows the maximum number of entries in a 
finger table (y-axis) against the number of device in the network (x-axis).
The graph shows a logarithmic growth with the number of devices and a linear 
pattern with the number of \emph{overlays}.
This is due to the design of the finger table 
and of $\text{SAFE}^d$.
This table, in fact, is used by Chord to optimize packets
routing (see Section~\ref{ssec:chord}) and contains at most 
a logarithmic number of entries with respect 
to the network size. Moreover, each node maintains a 
separate table for every \emph{overlay}. The overall
memory cost is then logarithmic in the number of devices
and linear in the number of \emph{overlays}.

\paragraph*{\textbf{Proofs Distribution}}
Figure~\ref{fig:random} shows an empirical analysis of the \emph{proofs} 
distribution in a simulated network with $10$K devices and using different 
\emph{overlays} (from $1$ to $3$).
On average, each device should contain a number of \emph{proofs} equal to the
number of \emph{overlays}.
However, since the distribution is random, some device may contain more 
\emph{proofs} than others.
The number of \emph{overlays} helps improve the \emph{proofs} distribution.
In fact, with $3$ \emph{overlays} the curve tends to be smoother.
In the worst scenario, we measured only less than $5$ devices that
contained $25$ \emph{proofs} in a network with $10$K elements and 
$3$ \emph{overlays}.
In case of $3$ \emph{overlays}, a network contains $30$K \emph{proofs}, $3$ for each device.
Our approach allows to build a collaborative attestation 
by solely requiring a device to contain $25$ \emph{proofs} 
at most, which makes $\text{SAFE}^d$ scalable.

To sum up, the number of proofs $p$ that a device stores is bounded and can be forced to reach the ideal value (\ie $1$ \emph{proofs} for each \emph{overlay}). 
Considering the overall memory consumption, $\text{SAFE}^d$ is a clear improvement with respect to previous works that are either 
an order of magnitude more expensive~\cite{pasta} or have a quadratic dependence on unpredictable parameters~\cite{us-aid}.


\subsection{Resilience}
\label{ssec:resilience}
The purpose of this experiment is to measure the ability 
of $\text{SAFE}^d$ to recover missing \emph{proofs} in case 
of attacks or faults.
In this scenario, we modeled a powerful attacker that randomly
destroys all the \emph{proofs} of a device.
In a real case, this simulates an adversary that 
physically destroys a device, interrupts the \emph{normal world}
scheduling, or simply a fault in the network.
This attack is tuned by a drop rate which indicates 
the number of devices that drops all their \emph{proofs} on every
simulation cycle.
We experimented a drop rate of $20\%$ with a simulation cycle of $10$ seconds, which means that on average $20\%$ of the 
entire network
erased all of its proofs every $10$ seconds.
We tested a different number of
\emph{overlays} to observe the different responses.
Figure~\ref{fig:resilience} shows the results of our experiments
in a simulated network that contains up to $10$K devices and different 
number of \emph{overlay} (from $1$ to $3$).
The y-axis shows the \emph{resilience index}, 
which is the ratio between the number of \emph{proofs}
correctly recovered and the number of lost \emph{proofs}.
The x-axis, instead, shows the number of devices in the network.
According to the attestation algorithm
(Section~\ref{ssec:attestation}), in case of lost \emph{proofs},
the \emph{Verifier} attempts to recover the
missing information from the other \emph{overlays}.
Therefore, the \emph{resilience index} tends to $1$ if all the \emph{proofs} were correctly recovered, otherwise it goes to $0$.
The plot shows that with the increases of the \emph{overlays}, the 
\emph{resilience index} tends to reach $1$ even in the presence 
of an high drop rate.
More precisely, we manage to recover $99.9\%$ of the \emph{proofs} with $3$ \emph{overlays}.
This experiment shows that the \emph{overlays} can be effectively used to recover the network in the presence of attacks.

\subsection{Security Consideration}
\label{ssec:securityconsideration}
We describe how $\text{SAFE}^d$ reacts against different attacker scenarios.

\paragraph*{\textbf{Tampered Devices}}
An attacker may infect a device and take control of it.
Since we use a trusted anchor, we consider the 
\emph{secure world} as protected, while the 
\emph{normal world} can be under attacker control.
Therefore, $\text{SAFE}^d$ protocol is protected by design.
Moreover, all the packets that transit through the 
\emph{normal world} are encrypted, thus outside 
the attacker range.
However, an attacker may avoid invoking trusted anchor code 
compromising \emph{normal world} scheduler. 
In this case, if the \emph{trusted world} is not triggered,
the Chord protocol cannot work properly, and the neighbour devices
can realize the attack.

To test $\text{SAFE}^d$ effectiveness, we verified that the other devices are able to spot the modified code inside the \emph{normal world}.

\paragraph*{\textbf{Attacks against the Network}}
All the messages exchanged among trusted anchors are encrypted
(Section~\ref{ssec:secure-device-comm})
and only devices that joined the network can communicate 
among each other
(Section~\ref{ssec:certification-phase}).
Man-in-the-middle~\cite{asokan2003man} attacks are 
mitigated by design:
\begin{enumerate*}[label=(\roman*)]
	\item the body is protected by the symmetric key $K_{AB}$,
	\item the header can be manipulated only by the trusted anchors of authorized devices.
\end{enumerate*}
We also include \emph{nonces} to avoid replay attacks.
This enhances robustness even in case of corrupted devices 
as long as their trusted anchor remains intact.

\paragraph*{\textbf{Physical attacks}}
According to DARPA attacker model~\cite{darpa}, a device 
which receives a physical attack is temporarily removed 
from the network.
Previous authors~\cite{darpa,scapi,pasta} proposed to use 
a heartbeat to keep the devices synchronized. In this way, 
a device that goes temporarily off-line cannot get
aligned with the heartbeat, and therefore, enables 
the detection of the attack.
However, establishing a heartbeat protocol implies the 
presence of loosely synchronized and secure clocks in 
every device. $\text{SAFE}^d$ overcomes this requirement 
by using the communication timeouts and \emph{nonces} 
to detect network disconnections.
During the protocol execution, each device periodically 
contacts its successor to assess its status.
In case a timeout occurs, the device uses the successor list 
to contact the closest node following the old one.
The contacted device will further check if its predecessor 
left the network, and if so, it will acquire the message 
sender as new predecessor, while launching an alert for 
a possible physical attack.
The double check adds robustness against 
simple network malfunctions.
To enhance the protection against physical attacks, we 
further propose a certificate revocation strategy that 
will be discussed in Section~\ref{sec:discussion}.

\paragraph*{\textbf{Denial-of-Service}} 
We do not protect against denial-of-service in case of 
a network entirely compromised 
(\eg all the messages are dropped). 
However, we can partially recover information loss by 
combining multiple \emph{overlays} and our attestation 
protocol (Section~\ref{ssec:resilience}).
\section{Discussion}
\label{sec:discussion}


\paragraph*{\textbf{Certificate Revocation}} 
$\text{SAFE}^d$ security properties can be further improved by adopting an efficient certificate revocation mechanism.
This feature can be useful in at least three scenarios:
\begin{enumerate*}[label=(\roman*)]
	\item if the CA private key gets compromised (\eg leaked),
	\item if a software is updated and
	\item if a device is corrupted.
\end{enumerate*}
The design of a scalable and efficient certificate revocation procedure was already addressed by~\cite{bloom1,bloom2} that proposed solutions based on Bloom filters~\cite{bloom3}.
Furthermore, ~\cite{bloom4} proposed a way to make a Bloom filter scalable, \ie to make its capacity adaptable at runtime so that it can be increased without stopping the general execution.
It is possible to implement in $\text{SAFE}^d$ a certificate revocation protocol that is scalable and distributed based on the previous citations.

\paragraph*{\textbf{Run-time Attacks}} An attacker could alter the application 
behavior without modifying the binary by using run-time
attacks~\cite{carlini2014rop}.
A way to cope with those threats is using run-time remote
attestation~\cite{DBLP:journals/corr/abs-1807-08003,abera2016c} that can verify
run-time properties, \eg the current execution path.
Usually, these solutions require several \emph{proofs} to be stored.
We can use the DHTs in $\text{SAFE}^d$ to spread the \emph{proof} load among devices.
We leave this as a future work.

\paragraph*{\textbf{Run-time Software Upgrade}} In specific cases (\eg industrial
IoT), we need to upgrade devices software without interrupting the network.
In $\text{SAFE}^d$, we do not deal with this case, but it is possible to mitigate this issue
by using two main approaches:
\begin{enumerate*}[label=(\roman*)]
	\item we could introduce new upgraded devices in the network and remove 
	the old ones until all the network is upgraded,
	\item we integrate specific upgrade protocols in $\text{SAFE}^d$ that load
	new software and substitute the \emph{proofs} in the DHTs.
\end{enumerate*}
Regardless the strategy adopted, the software upgrade strategy should be integrated with a strong certificate revocation mechanisms to avoid an attacker to re-upload old and vulnerable software.
\section{Conclusion}
\label{sec:conclusion}

In this work, we proposed $\text{SAFE}^d$, the first concrete
self-attestation schema for networks of heterogeneous embedded devices.
$\text{SAFE}^d$ maintains multiple copies of the \emph{proofs} among the devices,
which are equipped with small trusted anchors.
We also designed and developed new techniques that enhance classic DHT protocols 
against powerful adversaries, which are typical of remote attestation scenarios.

$\text{SAFE}^d$ allows performing remote attestations without 
the need of an external \emph{Verifier}, and consequently removing a single 
point of failure.
$\text{SAFE}^d$ coordinates multiple devices to self-protect the network and also to 
self-recover missing or corrupted \emph{proofs} in presence of attackers and 
faults.

We implemented a prototype of $\text{SAFE}^d$ in the open-source platform 
Raspberry Pi 3, this allows us to show the technical challenges 
faced for the implementation of $\text{SAFE}^d$ in the ARM 
TrustZone architecture.
Moreover, we stressed $\text{SAFE}^d$ performances by simulating a network of 
$10$K devices.
As a result, we showed that $\text{SAFE}^d$ requires a logarithmic amount
of memory and a logarithmic time to perform a complete attestation.
Moreover, $\text{SAFE}^d$ can recover up to $99.9\%$ of \emph{proofs} in case 
of attack or faults by using only three \emph{overlays}.


\bibliographystyle{IEEEtranS}
\bibliography{biblio}

\end{document}